\documentclass[aps,prb,twocolumn,superscriptaddress,showpacs]{revtex4}

\usepackage{graphicx}
\usepackage{dcolumn}
\usepackage{bm}
\usepackage{amsmath}
\usepackage{subfigure}

\newcommand{\bk}{\mathbf{k}}
\newcommand{\bq}{\mathbf{q}}

\begin{document}

\preprint{22241-LDA_v1}

\title{Self-doping effect and possible antiferromagnetism at titanium-layers in the iron-based superconductor Ba$_2$Ti$_2$Fe$_2$As$_4$O}

\author{Hao Jiang}
\affiliation{Department of Physics, State Key Lab of Silicon Materials, and Center for Correlated Matter,
Zhejiang University, Hangzhou 310027, China}

\author{Yun-Lei Sun}
\affiliation{Department of Physics, State Key Lab of Silicon Materials, and Center for Correlated Matter,
Zhejiang University, Hangzhou 310027, China}

\author{Jianhui Dai}
\affiliation{Condensed Matter Physics Group,
  Department of Physics, Hangzhou Normal University, Hangzhou 310036, China}

\author{Guang-Han Cao}
\email[E-mail address: ]{ghcao@zju.edu.cn}
\affiliation{Department of Physics, State Key Lab of Silicon Materials, and Center for Correlated Matter,
Zhejiang University, Hangzhou 310027, China}

\author{Chao Cao}
  \email[E-mail address: ]{ccao@hznu.edu.cn}
\affiliation{Condensed Matter Physics Group,
  Department of Physics, Hangzhou Normal University, Hangzhou 310036, China}

\date{\today}

\begin{abstract}
The electronic structure of Ba$_2$Ti$_2$Fe$_2$As$_4$O, a newly discovered superconductor, is investigated using first-principles calculations based on local density approximations. Multiple Fermi surface sheets originating from Ti-3$d$ and Fe-3$d$ states are present corresponding to the conducting Ti$_2$As$_2$O and Fe$_2$As$_2$ layers respectively. Compared with BaFe$_2$As$_2$, sizeable changes in the related Fermi surface sheets indicate significant electron transfer (about 0.12$e$) from Ti to Fe, which suppresses the stripe-like antiferromagnetism at the Fe sites and simultaneously induces superconductivity. Our calculations also suggest that an additional N\'{e}el-type antiferromagnetic instability at the Ti sites is relatively robust against the electron transfer, which accounts for the anomaly at 125 K in the superconducting Ba$_2$Ti$_2$Fe$_2$As$_4$O.
\end{abstract}

\pacs{74.70.Xa; 71.20.Ps; 75.30.Fv; 74.20.Pq}


\maketitle

The discovery of high-temperature superconductivity in iron pnictides\cite{hosono,cxh} has triggered stimulated studies in the condensed matter physics field.\cite{reviews} A common feature of these Fe-based superconductors in various families are that they all contain Fe$_2X_2$ ($X$=As or Se) layers in which the Fe moment aligns antiferromagnetically\cite{1111afm,122afm,11afm,245afm} at the ground state of the parent compounds. Superconductivity emerges if the antiferromagnetism (AFM) is sufficiently suppressed by either chemical doping\cite{hosono,K-doping,P-doping} or applying pressures.\cite{pressure} The electronic structure of the undoped Fe$_2X_2$ layers is semi-metallic, characterized by two or more hole pockets around the $\Gamma$ (0, 0) point, and two electron pockets near the $X$ ($\pi$, $\pi$) point.\cite{dh,terashima} First-principles studies based on density functional theory (DFT) have been very helpful in obtaining the band structure,\cite{singh2008} predicting the superconducting pairing symmetry,\cite{mazin2008} constructing the theoretical model,\cite{kuroki2008,zsc,cc2008} and reproducing the experimental magnetic ground states.\cite{mazin2008,11dft-mag,245dft-mag}

\begin{figure}
\includegraphics[width=7.5cm]{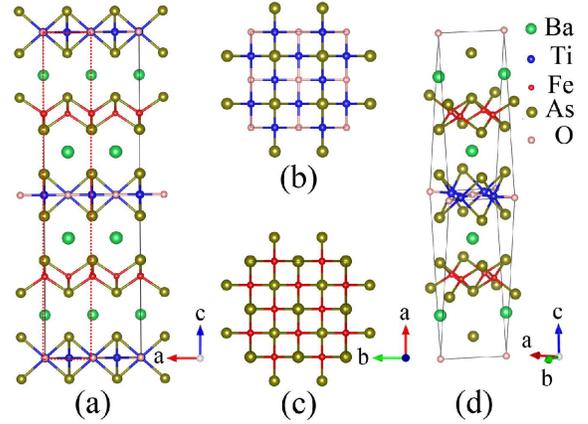}
\caption{Crystal structure of Ba$_2$Ti$_2$Fe$_2$As$_4$O superconductor. (a) a [100] side view showing the layered structure containing Ti$_2$O sheets as well as the common Fe$_2$As$_2$ layers. Top views on Ti$_2$As$_2$O sheets and Fe$_2$As$_2$ are displayed in (b) and (c). A primitive cell is shown in (d).\label{fig:structure}}
\end{figure}

By building new structures containing Fe$_2X_2$ layers, a new oxypnictide superconductor Ba$_2$Ti$_2$Fe$_2$As$_4$O (Ba22241) was recently discovered.\cite{22241} The Ba22241 material can be viewed as an intergrowth of BaFe$_2$As$_2$\cite{122} and BaTi$_2$As$_2$O\cite{1221}, containing Ti$_2$As$_2$O sheets as well as the common Fe$_2$As$_2$ layers (see Fig. \ref{fig:structure}). Although neither BaFe$_2$As$_2$ nor BaTi$_2$As$_2$O is superconducting, the combined structure--Ba22241--shows superconductivity at 21 K \emph{without doping}. Based on the variation of bond-valence sum (BVS) of Ti, electron transfer from Ti to Fe was proposed to interpret the appearance of superconductivity as a result of self doping.\cite{22241} Another interesting issue comes from an anomaly at $T^{*}\sim$125 K in electrical resistivity and magnetic susceptibility. The origin and the nature of this anomaly remain mysterious because both Fe and Ti sublattices could play a role. It was tentatively ascribed to a possible charge-density wave (CDW) or spin-density wave (SDW) transition in the Ti$_2$As$_2$O layers in the original study.\cite{22241} This proposal was supported by recent M\"{o}ssbauer experiment which indicates paramagnetic (PM) state, without long-range magnetic ordering below the $T^{*}$, for the Fe in Ba22241.\cite{mossbauer} So, the Ba22241 material may represent a rare example to show a two-dimensional (2D) CDW/SDW ordering sandwiched by superconducting layers.

To clarify and interpret the above experimental results, in this communication, we present a detailed first-principles study of the Ba22241 electronic structure by DFT calculations, in direct comparison with both BaFe$_2$As$_2$ and BaTi$_2$As$_2$O. It is shown that the density of states (DOS) around Fermi level $E_F$ are dominated by Fe-3$d$ orbitals in the FeAs layers as well as Ti-3$d$ and As-4$p$ orbitals from the Ti$_2$As$_2$O layers. The Fermi surface sheets can be regarded as a combination of BaFe$_2$As$_2$ and BaTi$_2$As$_2$O systems, respectively. Nevertheless, sizeable changes in the corresponding Fermi surface sheets, which means significant electron transfer to the FeAs layers, are induced by the insertion of Ti$_2$As$_2$O layers. The charge transfer serves as a self-doping effect, which suppresses the formation of long range magnetic ordering in the FeAs layers and leads to superconductivity without explicit doping. The calculated bare electron susceptibility suggests an AFM instability in the Ti$_2$As$_2$O layers, similar to that in BaTi$_2$As$_2$O. This AFM instability is itinerant in nature, but is weakened compared to that of BaTi$_2$As$_2$O due to the self-doping effect, thus it may account for the anomaly at 125 K in Ba22241.

All reported results are obtained using the DFT as implemented in the Vienna abinitio simulation package (VASP)\cite{vasp_1,vasp_2}, where the valence electron-ion interactions are modeled using projected augmented wave (PAW) method\cite{paw_bloch,vasp_2} and the exchange-correlation effects are approximated with Perdew-Burke-Enzerhoff flavor of general gradient approximation (PBE).\cite{PBE_1} To ensure the convergence of the calculations, a 520 eV energy cutoff to the plane wave basis and a 9$\times$9$\times$9 $\Gamma$-centered K-mesh is employed; whereas a $18\times18\times18$ $\Gamma$-centered K-mesh and tetrahedron method are employed to perform the DOS calculations.

Since the M\"{o}ssbauer experiment results show that long-range magnetic ordering is absent at the FeAs layers, most of our calculations are based on the spin-unpolarized DFT. In fact, due to the systematic overestimation of magnetic energies and moments in spin-polarized DFT calculations of iron-pnictides\cite{mazin_dft_considerations}, spin-unpolarized DFT calculations in most cases will yield more useful information about the electronic structure of these systems\cite{mazin_dft_considerations,singh_review}. It has also been shown that the electronic structure of iron-pnictides sensitively depends on the As-height $z_{\mathrm{As}}$, whereas spin-unpolarized DFT calculations of iron-pnictides systematically underestimate $z_{\mathrm{As}}$\cite{mazin_dft_considerations} as well as the crystal $c$-axis. Similar effect has been observed in our calculations, as both $z_{\mathrm{As}}^{\mathrm{I}}$ (As$^{\mathrm{I}}$ denotes the arsenic atoms close to Fe sheets) and $c$-axis are substantially smaller (over 9\%) than the experimental value after full structural optimization, resulting in a significantly increased As-Fe-As bond angle (120.4$^{\circ}$ as oppose to the experimental value 112.1$^{\circ}$). Thus the experimental structure was adopted to perform calculations unless otherwise specified.

\begin{figure}
\centering
  \subfigure{
   \includegraphics[width=7.5cm]{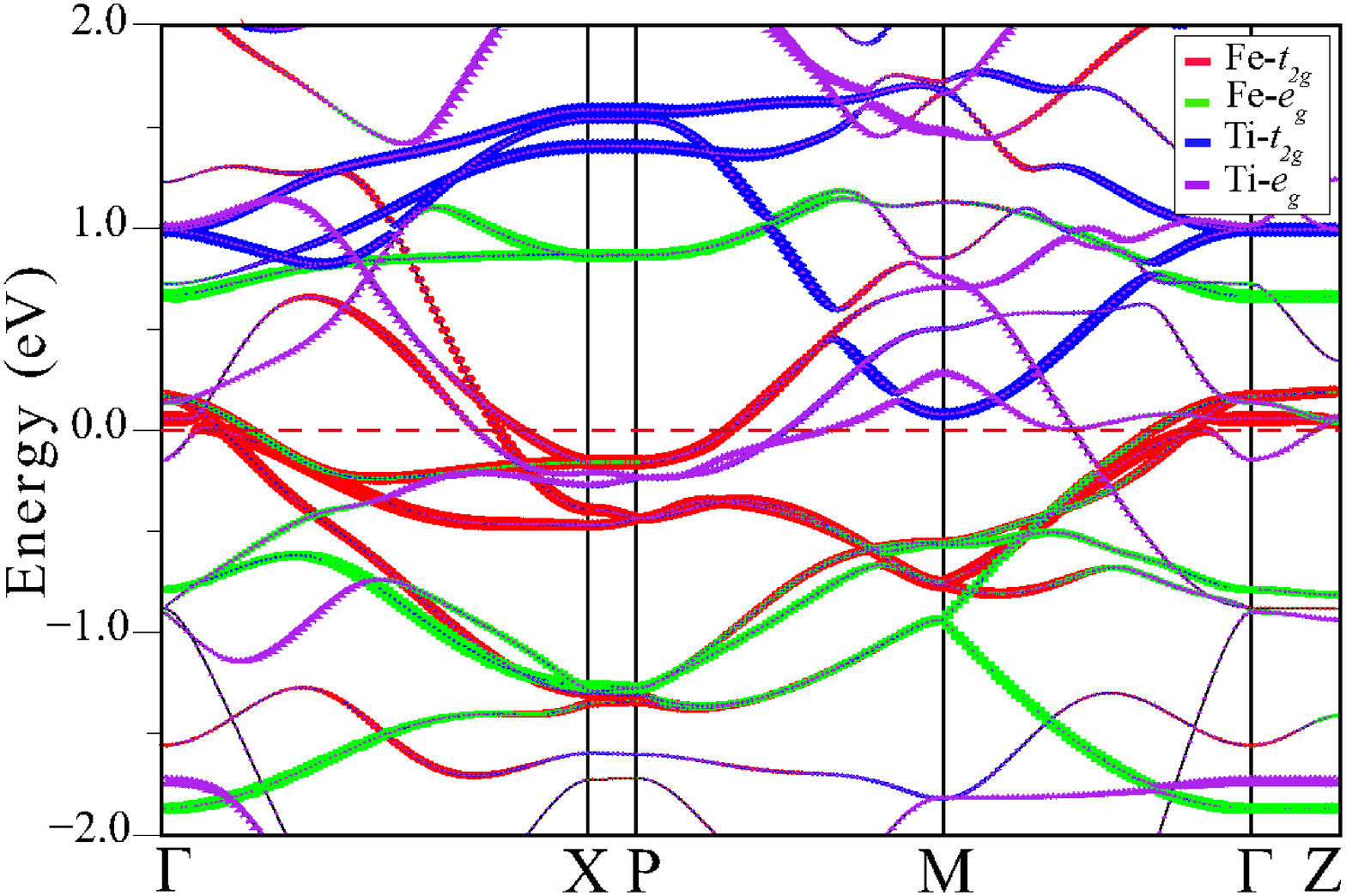}
  }
  \subfigure{
   \includegraphics[width=7.5cm]{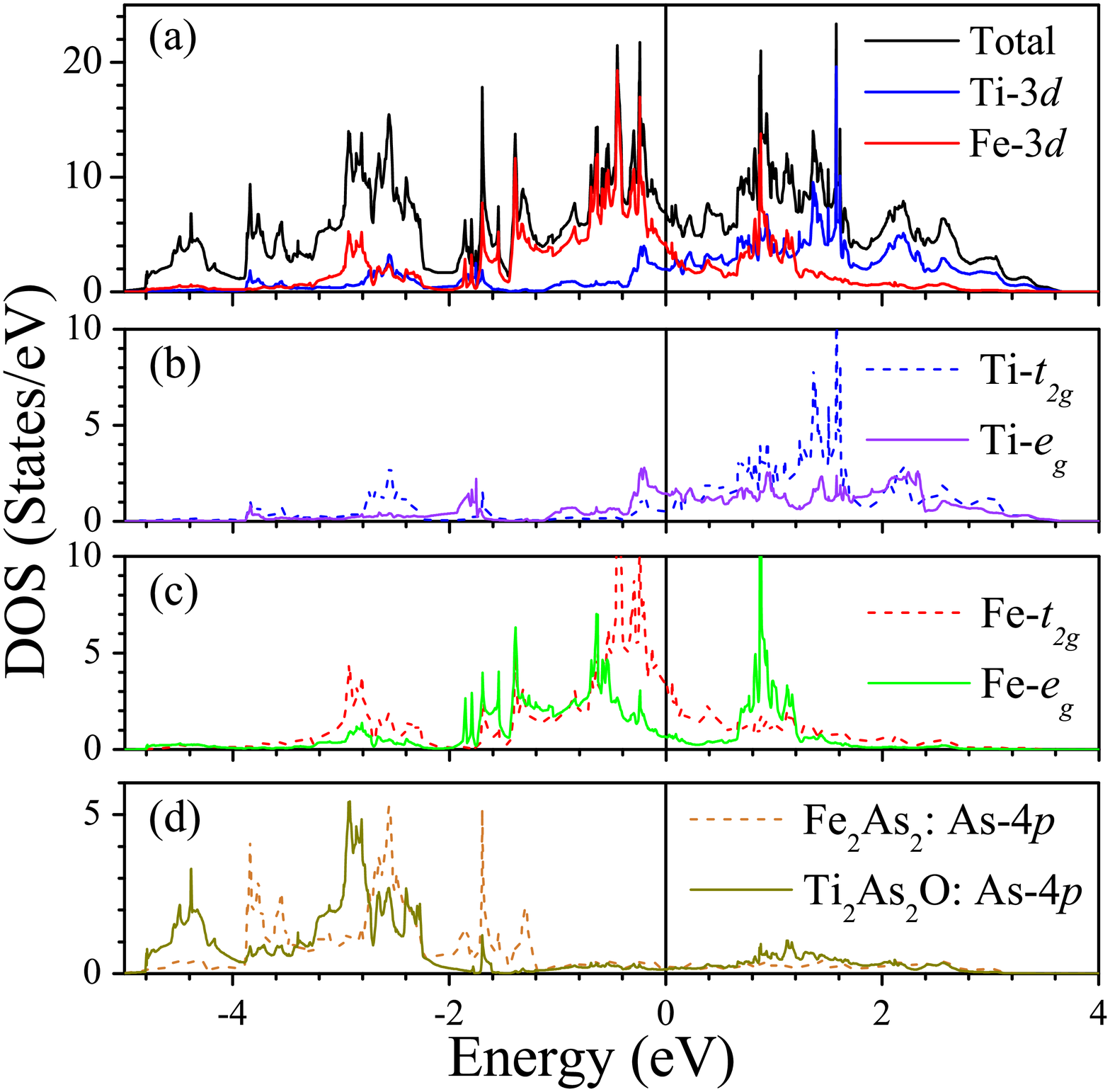}
  }
\caption{(Color online) Ba$_2$Ti$_2$Fe$_2$As$_4$O (a) band structures and (b) DOS as well as their projection onto different orbitals. The colors and widths of the lines in (a) represent the orbital character and its weight of the bands, respectively. In both figures, Fermi levels are shifted to zero.\label{fig:bsdos}}
\end{figure}

We show the band structure and density of states (DOS) of the nonmagnetic (NM) Ba22241 in Fig. \ref{fig:bsdos}. The Fe-3$d$ and Ti-3$d$ orbitals dominate the electronic states near the Fermi level $E_F$, contributing 38\% and 51\% of the DOS, respectively. The band dispersion within $E_F\pm0.5$ eV could be regarded as a simple superposition of BaFe$_2$As$_2$ band structure and the BaTi$_2$As$_2$O one. Close to $E_F$, the bands are dominantly Fe-$t_{2g}$ and Ti-$e_{g}$ orbitals, although the orbital character weight of the latter is significantly reduced due to the hybridization between Ti-3$d$ and As$^{\mathrm{I}}$-4$p$. Interestingly, all the Fe-related bands show little dispersion along $\Gamma-Z$ while most Ti-related bands disperse strongly along that direction, meaning that the FeAs layers are strongly two-dimensional while the Ti$_2$As$_2$O-layers are much more isotropic due to more itinerancy for the Ti-3$d$ electrons.

\begin{figure}
\centering
  \subfigure{
   \includegraphics[width=7.5cm]{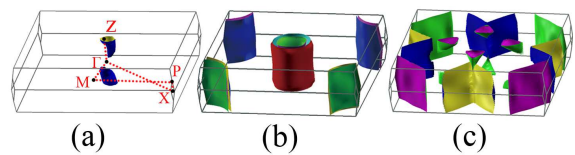}
  }
\caption{(Color online) (a) High symmetry points of the body-centered tetragonal primitive cell and the small Fe-related hole pocket around $\Gamma$. (b) Fe-related and (c) Ti-related Fermi surface sheets of Ba$_2$Ti$_2$Fe$_2$As$_4$O. Notice the shape of Fe-related sheets are very close to those of BaFe$_2$As$_2$. \label{fig:fs}}
\end{figure}

The NM Fermi surfaces were then obtained by fitting the DFT band structure to a tight-binding Hamiltonian using the maximally localized Wannier functions (MLWFs)\cite{mlwf_1,mlwf_2,fs_fit_detail}, and interpolating to a dense $100\times100\times100$ $\Gamma$-centered K-grid (Fig. \ref{fig:fs}). We separate the Ti-related and the Fe-related Fermi surface sheets in Fig. \ref{fig:fs}. Consistent with the band structure, the Fe-related hole pockets show less dispersion along $k_z$ axis than those of BaFe$_2$As$_2$ compound, suggesting an enhanced quasi-2D feature of the FeAs layer. To analysis the charge transfer between FeAs layers and Ti$_2$As$_2$O layers, we integrated the volume enclosed by the Fe-related hole-sheets around $\Gamma$. Parent BaFe$_2$As$_2$ has three hole sheets around $\Gamma$, which is integrated to be 0.335 $\vert e\vert$. For Ba22241, there are also three hole sheets, but the enclosed volume is reduced to 0.215$\vert e\vert$. Thus, $\sim$ 0.12 electron per unit cell is transferred to the FeAs layer from the Ti$_2$As$_2$O layer. This result is qualitatively consistent with both the projected-DOS analysis and an extra Bader charge analysis, which shows charge transfer by $\sim$ 0.05 per unit cell to the FeAs layer. We notice that the quantity of transferred charge is also in good agreement with the BVS analysis.\cite{22241} Previous studies\cite{reviews} have established that doping will suppress the formation of long-range stripe-like (SL-AFM$^{\mathrm{Fe}}$) order and thus induce the superconductivity. Therefore, we conclude that superconductivity in the undoped Ba22241 is induced by self doping due to the electron transfer from the intercalated Ti$_2$As$_2$O layer. It is noted that similar self-doping induced superconductivity was evidenced in Sr$_2$VFeAsO$_3$ superconductor.\cite{cao2010}

\begin{figure}
\centering
  \subfigure[Ba22241]{
   \rotatebox{270}{\scalebox{0.5}{\includegraphics{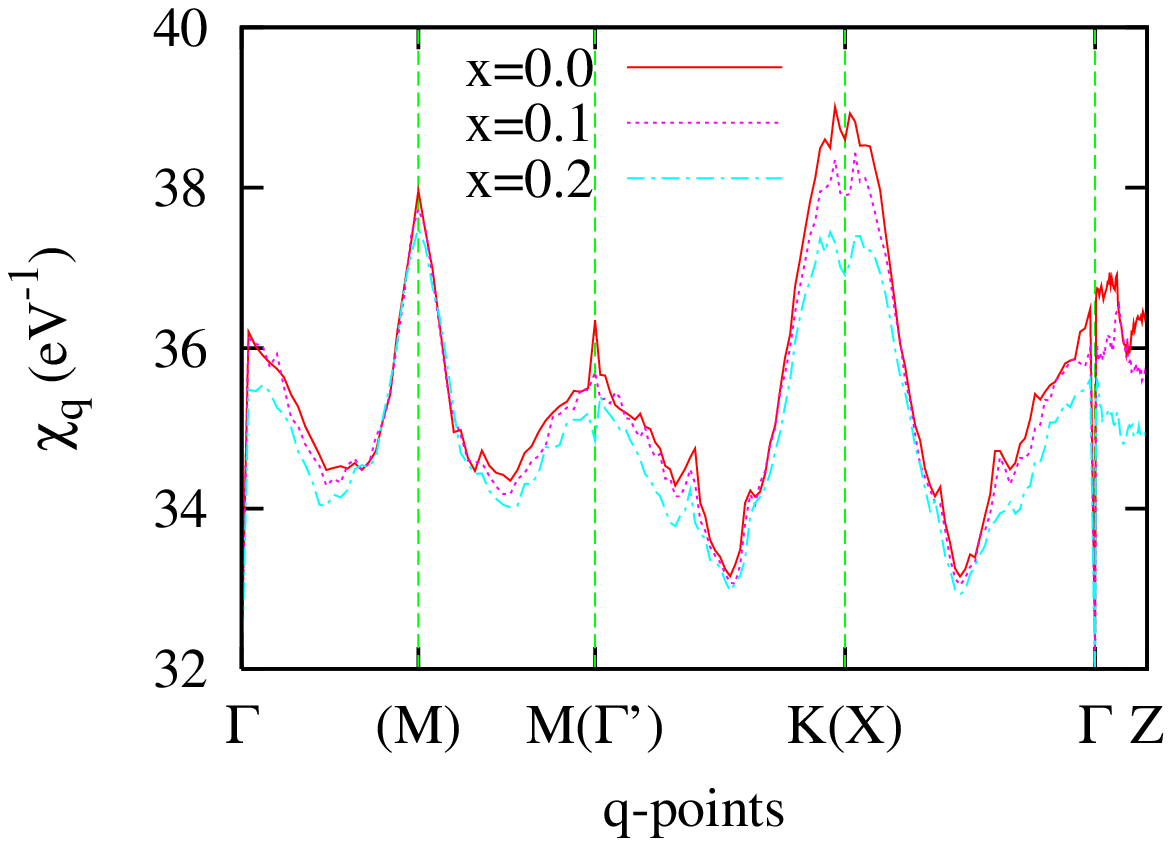}}}
  }
  \subfigure[Ba1221]{
   \rotatebox{270}{\scalebox{0.5}{\includegraphics{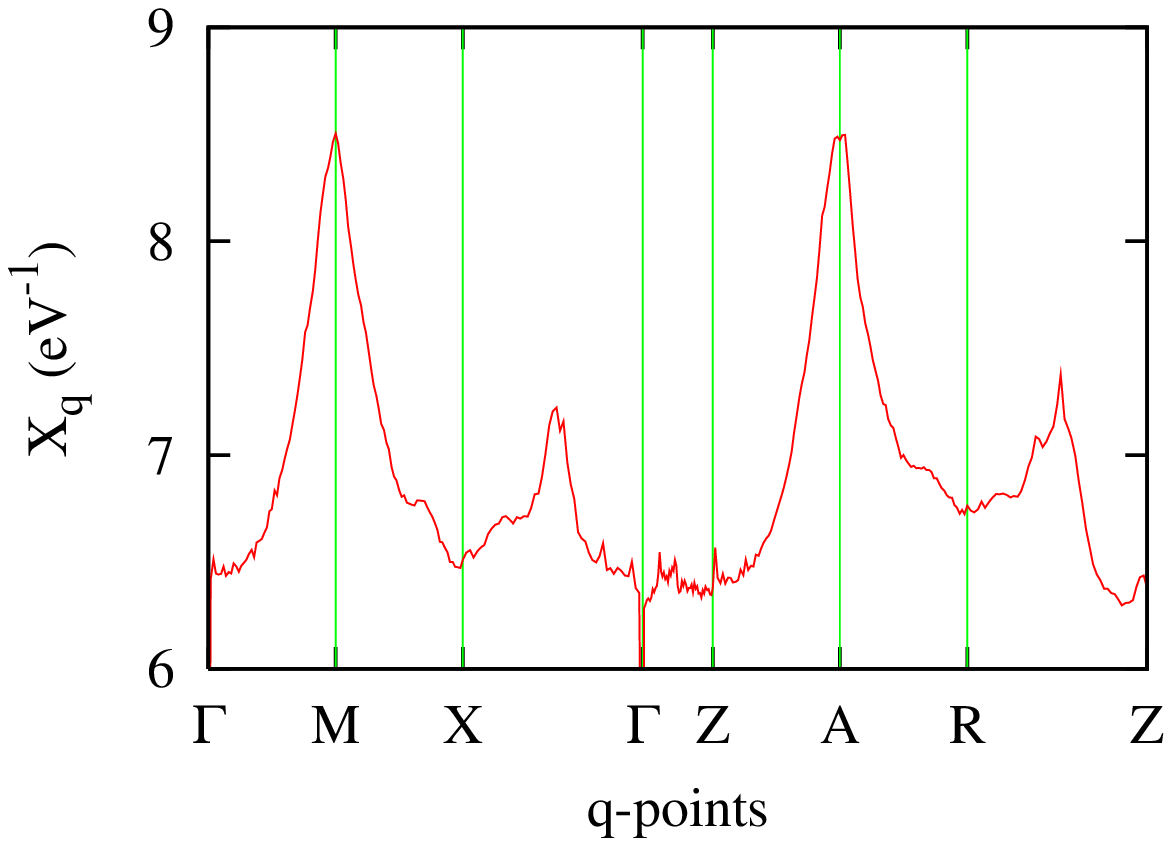}}}
  }
  \subfigure[Ba122]{
   \rotatebox{270}{\scalebox{0.5}{\includegraphics{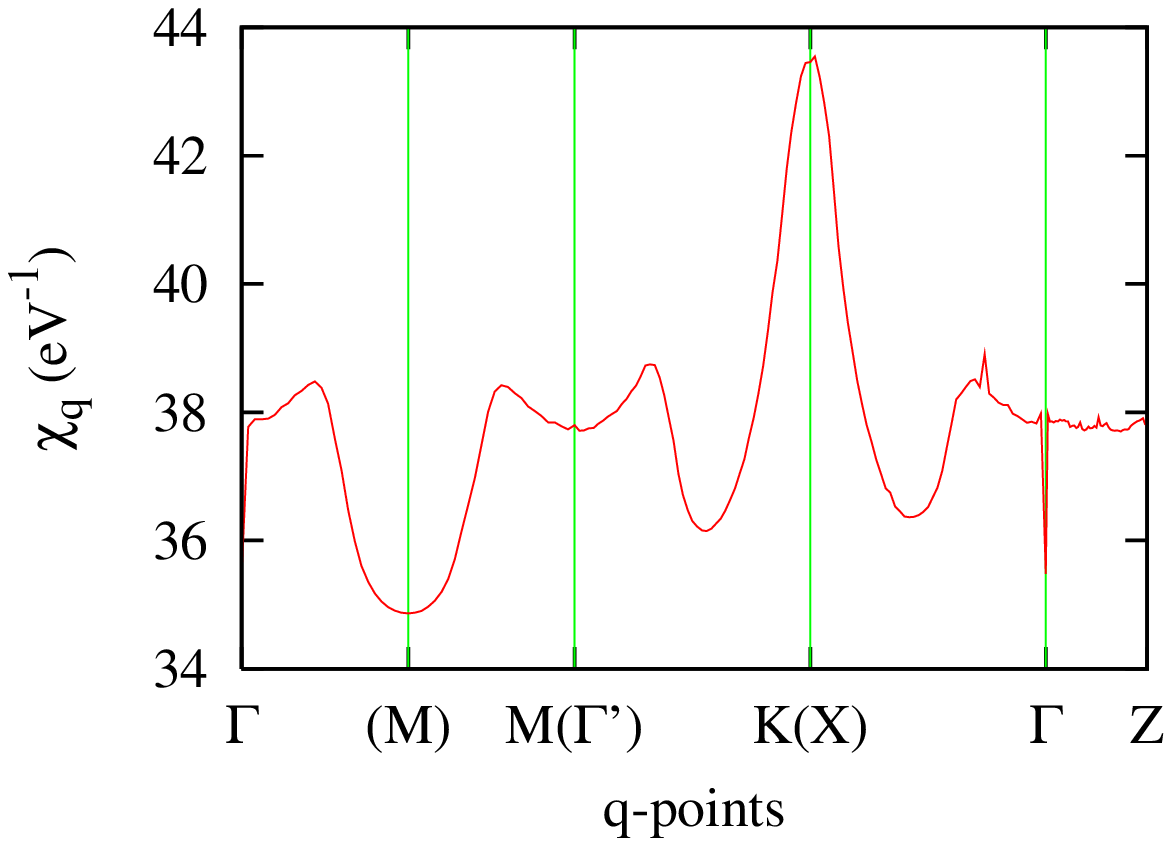}}}
  }
\caption{(Color online) Bare spin susceptibility of (a) Ba$_2$Ti$_2$Fe$_2$As$_4$O, (b) BaTi$_2$As$_2$O, and (c) BaFe$_2$As$_2$. In panel (a), the red solid line, blue dotted line and cyan dash-dotted line are for without doping, 0.1$e$ doping and 0.2$e$ doping, respectively. For Ba$_2$Ti$_2$Fe$_2$As$_4$O and BaFe$_2$As$_2$, the high symmetry points are labeled according to the Brillouin zone of the body-centered tetragonal primitive cell, the corresponding high symmetry points in the BZ of the tetragonal conventional cell are labeled inside the parenthesis. \label{fig:chi}}
\end{figure}

It is expected that the long range SL-AFM$^{\mathrm{Fe}}$ order is suppressed due to the self-doping effect, as also evidenced by the M\"{o}ssbauer experiment\cite{mossbauer}. However, this does not imply the absence of magnetic order in the Ti$_2$As$_2$O-layers. Itinerant magnetic instability may develop due to the metallic band structure of that layer. Therefore, we calculate the bare irreducible susceptibility using the maximally localized Wannier functions (MLWFs)\cite{mlwf_1,mlwf_2} according to
$$\chi_0(\mathbf{q})=\sum_{nm\bk}\frac{f(\epsilon_{n\bk})-f(\epsilon_{m\bk+\bq})}{\epsilon_{m\bk+\bq}-\epsilon_{n\bk}}$$
for Ba22241, BaTi$_2$As$_2$O, and BaFe$_2$As$_2$ compounds, as shown in Fig. \ref{fig:chi}. It should be noted that the plots for Ba22241 and BaFe$_2$As$_2$ are generated using primitive cells, and thus the corresponding Brillioun zone (BZ) needs to be folded to compare with conventional notations and with the BaTi$_2$As$_2$O plot. Originally, the formation of SL-AFM$^{\mathrm{Fe}}$ order in BaFe$_2$As$_2$ is signaled by a large peak at $X$ ($\pi$, $\pi$) in the folded tetragonal BZ\cite{notation_convention}. For Ba22241, the ($\pi$, $\pi$) peak is appreciably broadened, with a cusp forming at its center. It suggests that the original SL-AFM$^{\mathrm{Fe}}$ order is becoming unstable in Ba22241, consistent with the self-doping scenario. In addition to this feature, another strong sharp peak at $M$ ($\pi$, 0) (notation in the conventional tetragonal BZ) can be identified in Ba22241. This signals certain electronic instabilities. Using the rigid band model, we have also obtained $\chi_0$ of Ba22241 at different doping concentrations ($x$=0.0, 0.1 and 0.2) [Fig. \ref{fig:chi}(a)]. The ($\pi$, $\pi$) peak corresponding to the SL-AFM$^{\mathrm{Fe}}$ ordering formation is quickly suppressed with electron doping, with a deeper cusp at its center. On the contrary, the ($\pi$, 0) peak is almost unaffected by the doping. Therefore, the ($\pi$, 0) instability is in fact more robust than the SL-AFM$^{\mathrm{Fe}}$ ordering, and it is most likely to be responsible for the 125-K anomaly.

Interestingly, the ($\pi$, 0) peak is also present in BaTi$_2$As$_2$O, whose susceptibility exhibits significant enhancement around $M$ and $A$, suggesting that the instability is related with the Ti$_2$As$_2$O-layers. It is worthy noting here that the BaTi$_2$As$_2$O compound has a similar anomaly in the resistivity at $\sim 200$K\cite{1221}. Thus, it is instructive to perform calculations for BaTi$_2$As$_2$O compounds. By using PBE0 hybrid functional, we confirm that a checker-board type configuration (N\'{e}el-AFM$^{\mathrm{Ti}}$) with $\sim0.12\mu_B$/Ti lowers the total energy by $\sim$13 meV/Ti, which is energetically close to the $\sim$ 200 K anomaly. As discussed previously, the electronic structure of Ba22241 seems to be well separated into FeAs-related and Ti$_2$As$_2$O-related, except for the charge transfer that causes self-doping effect, the N\'{e}el-AFM$^{\mathrm{Ti}}$ at Ti$_2$As$_2$O layers should also be present in Ba22241. Therefore, the 125K anomaly in Ba22241 is possibly due to a depressed N\'{e}el-AFM$^{\mathrm{Ti}}$ magnetism at the Ti$_2$As$_2$O-layers with small moment. As the formation of magnetic ordering will generally reduce the carrier density, a reduction in the resistivity is expected if the carrier mobility is not enhanced. The scenario is in consistency with the experimental observation.

In conclusion, we have performed electronic structure calculations on the newly-discovered superconductor Ba$_2$Ti$_2$Fe$_2$As$_4$O. The presence of intercalated Ti$_2$As$_2$O layers induces 0.12$e$ charge transfer to FeAs layers, suppressing the formation of the long-range SL-AFM$^{\mathrm{Fe}}$ order at the FeAs layer, and causes the 21-K superconductivity without explicit doping. Bare susceptibility $\chi_0$ exhibits a strong enhancement around $M$ point, which is also present in BaTi$_2$As$_2$O compound, suggesting a possible N\'{e}el-AFM$^{\mathrm{Ti}}$ ordering at the Ti$_2$As$_2$O layers in Ba22241. Therefore, the actual ground state should be a weak AFM in the Ti sublattice and superconducting in the FeAs layers. The coexistence of superconductivity and weak AFM, which develop in the FeAs and Ti$_2$As$_2$O layers separately, is due to the almost disentangled electronic structures of the two layers near the Fermi level. How the Ti-AFM affects the superconductivity in FeAs layers, and \emph{vice versa}, seem to be quite interesting, and calls for further studies.

\begin{acknowledgments}
This work is supported by NSFC under grant No. 11190023 and 10934005, the National Basic Research Program of China (No. 2010CB923003), as well as NSF of Zhejiang Province No. LR12A04003 and No. Z6110033. All calculations were performed on the High Performance Computing Center of Hangzhou Normal University.
\end{acknowledgments}



\begin{thebibliography}{00}
\bibitem{fs_fit_detail} The Ba22241 band structure is fitted using 35 orbitals including Fe-3d, Ti-3d, As-4p and O-2p; BaFe$_2$As$_2$ band structure is fitted using 10 Fe-3d orbitals; BaTi$_2$As$_2$O band structure is fitted using 16 orbitals including Ti-3d and As-4p.
\bibitem{notation_convention} The high symmetry points and all $\mathbf{q}$-vectors are corresponding to 2-Fe BZ throughout this paper. Therefore ($\pi$, 0) is the N\'{e}el-type AFM, while ($\pi$, $\pi$) is the SL-AFM.
\end{thebibliography}
\end{document}